\documentclass[
  aps,
  prd,
  reprint,
  superscriptaddress
]{revtex4-2}

\usepackage{graphicx}
\usepackage{dcolumn}
\usepackage{bm}
\usepackage{graphicx,amsmath,amssymb,amsthm}
\usepackage{xcolor}
\usepackage[normalem]{ulem}
\usepackage{thmtools, thm-restate}
\usepackage{epsfig}
\usepackage{epstopdf}
\usepackage{latexsym}
\usepackage{graphicx}
\usepackage{booktabs}
\usepackage{bbm}
\usepackage{color}
\usepackage{physics}
\usepackage{tensor}
\usepackage{verbatim}
\usepackage[caption=false]{subfig}
\usepackage{tikz}
\usepackage{bigints}
\usepackage{ifthen}
\usepackage{soul}
\usepackage[hidelinks]{hyperref}

\newcommand{\Mpl}{M_{\rm Pl}}
\newcommand{\mkk}{m_{\rm KK}}

\begin{document}

\preprint{APS/123-QED}

\title{A Universal Kaluza-Klein Graviton Cascade Rate}

\author{Vincent S. H. Lee}
\affiliation{Department of Physics, University of California, Berkeley, Berkeley, California 94720, USA}
\affiliation{Department of Physics, University of California, San Diego, La Jolla, California 92093, USA}

\author{Lisa Randall}
\author{Marcos Riojas}
\affiliation{Department of Physics, Harvard University, 17 Oxford Street, Cambridge, Massachusetts 02138, USA}

\date{\today}

\begin{abstract}
We obtain a general decay rate between Kaluza-Klein (KK) gravitons, resolving an ambiguity in the literature concerning these decays. The amplitude is determined not only by the mass $m_n$ of the decaying particle, but also by the momentum $q$ of the decay products in the parent's rest frame, and is suppressed relative to the expression commonly used in the literature by  $(q/m_n)^2$. Phase space gives a general $q^5$-dependent decay rate obtained earlier in Randall-Sundrum models.  We identify the cancellation that gives this result as a consequence of the massless 5D graviton,  and  argue for its universality even with scalar-induced KK violation by absorbing non-minimally coupled scalar fields into the warp-factor. This has significant implications for the potential for cascading dark matter models, and can be used to rule out existing Dark Dimension models, because the rapid cascade that was needed to evade constraints is severely suppressed. The amplitude is established by an explicit calculation that accounts for the backreaction of fields and branes and  applies broadly to non-minimal diffeomorphism invariant two-derivative couplings between scalar fields and gravity. 

\end{abstract}

\maketitle

\section{\label{sec:Intro} Introduction}

If Kaluza-Klein (KK) number \cite{Kaluza:1921tu,Klein:1926tv} is broken, heavier KK modes can decay into lighter ones. This was noted long ago in \cite{Kolb:1983fm}, see also \cite{Starkman:2000dy,Mohapatra:2003ah}. Separately, Dienes and Thomas \cite{Dienes:2011ja,Dienes:2011sa} suggested that dark matter could be multi-component, where individual components can decay into standard model particles in a way consistent with observations. For instance scalar fields, including an axion, might reproduce the necessary distributions and decays. 

More recently, Gonzalo, Montero, Obied, and Vafa (GMOV) \cite{Gonzalo:2022jac,Vafa:2024fpx} suggested this idea applies to KK gravitons in a large flat extra dimension. 
There KK number violation is assumed to arise through a scalar and depends on the model, but the spectrum of KK modes and the consequent kinematics is determined by the geometry.  
The merit to such a scenario is the relative model-independence. In this letter, we will argue that this advantage cuts both ways because it constrains the viability of any such model. In particular the decay rate down the tower must be sufficiently rapid, as an efficient cascade is needed to evade constraints at recombination \cite{Slatyer:2016qyl,Gonzalo:2022jac,Obied:2023clp,Law-Smith:2023czn}. 

Based on dimensional reasoning, averaging over polarizations, with $\vec{q}$ the decay product momentum in the parent's rest frame,
a decay of mass $m_n$ would  contribute
\begin{equation}
    \Gamma_{n \rightarrow m, l}=\tfrac{|\vec{q}|}{8 \pi m_n^2}\left|\mathcal{M}_{nml}\right|^2\sim\tfrac{|\vec{q}|}{8 \pi m_n^2}\left|\chi_{n m l}\right|^2 \left( \tfrac{M^4}{\Mpl^2}\right),
    \label{eq:channelrate}
\end{equation}
to the cascade rate, where $\chi_{nml}$ is the integrated overlap between the KK modes and $|\mathcal{M}_{nml}|^2$ is the squared amplitude for the decay channel, see Eq.~\eqref{eq:chidef}. 

In Eq.~\eqref{eq:channelrate} we have assumed Planck scale suppression, but in a warped extra dimension this can be enhanced. The placeholder quantity $M$, a scale with units of mass, determines the amplitude. Many estimates \cite{Mohapatra:2003ah,Gonzalo:2022jac,Obied:2023clp,Law-Smith:2023czn,Vafa:2024fpx} take $M=m_n$, even though purely on dimensional grounds $M=q$ is also a possibility. Because an efficient cascade is needed for most models to be viable, it is crucial to determine which scaling is correct.

An enhanced rate was also found by \cite{Giudice:2017fmj}, in disagreement with \cite{deGiorgi:2021xvm,Bonifacio:2019ioc}, see also \cite{Chivukula:2024nzt,Chivukula:2025pmk,Donini:2025qrf,Im:2024kuw,deGiorgi:2026qjp}. In this letter, we establish that the universal answer to this  question is that $M=q$ independent of the model. It is shown in App.~\ref{app:action} and App.~\ref{app:weyl} that any scalar responsible for KK violation can be absorbed in a warp factor.  We find the rate given for the Randall-Sundrum model (RS) \cite{Randall:1999ee,Randall:1999vf} in \cite{deGiorgi:2021xvm} applies universally to all such models, including flat extra dimensions, and scales as $(q/m_n)^5 m_n^3/\Mpl^2$ when decaying to two KK gravitons of comparable mass \footnote{The full formula for unequal masses is similarly threshold suppressed}. Intuitively,  small KK number violation is needed to avoid excessive loss of mass to (e.g.) kinetic energy of decay products, and furthermore to satisfy phenomenological constraints $n$ will typically be large at production, and so the suppression  of $(q/m_n)^4\sim(1/n)^2$ relative to the incorrect rate is enormous for viable model parameters. Most such models are immediately ruled out. 

We begin with a simple symmetry argument. Scalar induced  KK number violation is assumed in  Ref.~\cite{Gonzalo:2022jac} to give the cubic vertex $\frac{g_n}{M_{\mathrm{Pl}}} h_{(k)}^{* \mu \nu} \partial_\sigma h_{\nu \rho}^{(i)} \partial^\sigma h^{(j) \rho}{ }_\mu$, following the reasoning given in \cite{Mohapatra:2003ah}. It is important to retain the underlying tensor structure which they left suppressed, and to include every term that enters the cubic vertex. This is because these terms can cancel (and they do). They implicitly took these two derivatives to contract with each other, which gives a product of momenta 
\begin{equation}
    2 p_i \cdot p_j=m_i^2+m_j^2-m_k^2.
    \label{eq:pdotp}
\end{equation}
The symmetrized sum over external legs scales as $m_n^2$, where $m_n$ is the mass of the parent, so
$\mathcal{M}\sim\beta\,m_n^2/M_{\mathrm{Pl}}$ which is $\mathcal{O}(m_n^2)$ at threshold. Summed over decay channels this gives the total rate of Ref.~\cite{Gonzalo:2022jac}, Eq.~\eqref{eq:wrongdecayrate} below. This would have given $M=m_n$, but such terms cancel. 

The general argument for the suppressed decay rates can be most simply understood as follows. In 5D, with background metric $d s^2=e^{-2A(y)}\eta_{\mu\nu}dx^\mu dx^\nu+dy^2$, the graviton fluctuation is $H_{\mu\nu}(x,y)=\sum_i H^{(i)}_{\mu\nu}$ with $H^{(i)}_{\mu\nu}=h^{(i)}_{\mu\nu}(x)\psi_i(y)$. For two contracting derivatives connecting three external legs,  general covariance requires
\begin{equation}
\begin{aligned}
&\sqrt{- G}\, G^{AB}H^{(k)} \partial_A H^{(i)}\partial_B H^{(j)}\\
&\quad= H^{(k)} e^{-2A}\partial_\mu H^{(i)}\partial^\mu H^{(j)}
+ H^{(k)} e^{-4A}\partial_y H^{(i)}\partial_y H^{(j)}.
\end{aligned}
\label{eq:contract5D}
\end{equation}
Integrating by parts,  symmetry of the external legs implies (equivalently, summing over indices in the Lagrangian) that the contribution to the cubic-order gravitational action is proportional to
\begin{equation}
\int d^4 x \, dy \, H^{(j)}H^{(k)}\Big[e^{-2A}\Box_4 H^{(i)}
+\partial_y\big(e^{-4A}\partial_y H^{(i)}\big)\Big].
\label{eq:eomcancel}
\end{equation}
The bracket is just $\sqrt{- G}\,\Box_5 H^{(i)}$, which vanishes for any $A(y)$ by the equation of motion for a massless 5D graviton. This rule (no derivatives contracting together) descends from 5D to 4D through the mode expansion, and this cancellation explains why $M \ne m_n$. We verify this explicitly in Sec.~\ref{sec:cancel}. The surviving terms, which contract momenta with polarizations, scale with $q$ instead of $m_n$. This requires few assumptions and applies broadly, for example to the flat extra dimensions considered in \cite{Gonzalo:2022jac}.

We show this result holds independently of how KK number is broken by absorbing the fields responsible for KK number violation into a warp factor. Non-minimal couplings are similarly absorbed through a Weyl rescaling of the brane and bulk actions, giving a universal decay rate. Scaling relations for these decays are also found: the dominant mode in the tower satisfies $m(t) \propto t^{-2/3}$, rather than $m(t)\propto t^{-2/7}$ in \cite{Gonzalo:2022jac,Vafa:2024fpx}. Mass lost by the tower during these cascades also satisfies scaling relations given below. In particular, using the corrected decay rate and scaling relations, the suggestion by GMOV \cite{Gonzalo:2022jac,Vafa:2024fpx} is ruled out by decays to the Standard Model. 

In this letter, we will refer to this larger class of theories as ``Cascading Graviton Dark Matter" (CGDM), and consider in a companion paper \cite{Lee:toappear} whether one particular type of warping leaves room for viable parameter space, even with the heavily suppressed decay rate.

\section{The Cancellation for General Warp Factor}
\label{sec:cancel}

We show in an Appendix that scalars can be absorbed into a warp factor, as alluded to above, through Einstein's equations and a Weyl rescaling of non-minimally coupled scalars.  This is perhaps unsurprising in that warped compactifications, for instance the RS model \cite{Randall:1999ee,Randall:1999vf}, break KK number automatically, as in \cite{Cesarotti:2020uod}.

So consider the general warped ansatz, for which a flat extra dimension \cite{ArkaniHamed:1998nn} as in GMOV \cite{Gonzalo:2022jac,Vafa:2024fpx} is a special case
\begin{equation}\label{eq:ansatz}
    d s^2=\hat{g}_{\mu \nu}(x, y) d x^\mu d x^\nu+d y^2, \quad \hat{g}_{\mu \nu}=e^{-2 A(y)} g_{\mu \nu},
\end{equation}
where the warp factor can follow the background stress energy tensor or be sourced by a bulk scalar profile $\bar{\Phi}(y)$. This field, and jumps in $A'(y)$ from bulk branes, can break $y$-translation invariance, enabling the cascade.

The 5D Einstein-Hilbert term may be written, using the Gauss-Codazzi equation, in terms of the curvature of the slices, the extrinsic curvature, and a total derivative:
\begin{equation}
    \begin{aligned}
\sqrt{-G} R_5=\sqrt{-\hat{g}}(R[\hat{g}] & \left.+\hat{K}^2-\hat{K}^\mu{ }_\nu \hat{K}^\nu{ }_\mu\right) \\
& -2 \partial_y(\sqrt{-\hat{g}} \hat{K}),
\end{aligned}
\end{equation}
where $\hat{K}_{\mu \nu} \equiv \frac{1}{2} \partial_y \hat{g}_{\mu \nu}$ is the extrinsic curvature of the slices, and $\hat{g}$ denotes the induced metric on the slice. If there are no boundaries in this problem, then the total derivative vanishes. Otherwise the total derivative cancels against the Gibbons-Hawking-York (GHY) term.

The branes source jumps in the extrinsic curvature through the Israel junction conditions (IJCs):
\begin{equation}
    \left[\hat{K}_{\mu \nu}\right]=-\kappa_5^2\left(T_{\mu \nu}-\frac{1}{3} \hat{g}_{\mu \nu} T\right), \quad \kappa_5^2 \equiv M_5^{-3},
\end{equation}
with the brane stress tensors given in Appendix~\ref{app:action}. The extrinsic curvature on the slice satisfies
\begin{equation}
    \hat{K}^\mu{ }_\nu=-A^{\prime}(y) \delta^\mu{ }_\nu+K^\mu{ }_\nu, \quad K^\mu{ }_\nu \equiv \frac{1}{2} g^{\mu \rho} \partial_y g_{\rho \nu} ,
\end{equation}
where $K^\mu{}_\nu$ is the usual extrinsic curvature, built from $g$, and gives the term $K^2-K^\mu{}_\nu K^\nu{}_\mu$ in Eq.~\eqref{eq:genaction} below. The background is fixed by Einstein's equations. On maximally symmetric slices, the $(\mu \nu)$ equations on the slice have an on-shell matter Lagrangian where 
\begin{equation}
    \mathcal{L}_M=-\frac{M_5^3}{2}\left(6 A^{\prime \prime}-12 A^{\prime 2}\right).
    \label{eq:genmatteraction}
\end{equation}
This cancels corresponding bulk terms in $\hat{K}^2-\hat{K}^\mu{ }_\nu \hat{K}^\nu{ }_\mu$. The remaining boundary terms cancel exactly with the actions for the branes, which are determined by the IJC.

The surviving gravitational action then takes the form 
\begin{equation}
\frac{M_5^3}{2} \int d^4 x\, d y  \sqrt{-g} \left[ e^{-2 A} R[g] + e^{-4 A}\left(K^2-K^\mu{ }_\nu K^\nu{ }_\mu\right) \right],
\label{eq:genaction}
\end{equation}
where the extrinsic curvature term is:
\begin{equation}
    K^2-K^\mu{}_\nu K^\nu{}_\mu
=\tfrac14\big(g^{\mu\nu}g^{\rho\sigma}-g^{\mu\rho}g^{\nu\sigma}\big)
\partial_yg_{\mu\nu}\,\partial_yg_{\rho\sigma}.
\end{equation}

We find that non-minimal couplings give the same result, because the difference can also be absorbed into the warp factor $A(y)$. See App.~\ref{app:weyl} for a detailed argument. Briefly, the most general diffeomorphism invariant two-derivative action, $F(\tilde\Phi)\tilde R-\tfrac12 k(\tilde\Phi)(\partial\tilde\Phi)^2-\tilde V(\tilde\Phi)$ in the bulk with brane tensions $\tilde\lambda_b(\tilde\Phi)$ and the GHY term $2\int\sqrt{-\tilde g_{\rm ind}}\,F\tilde K$, becomes minimally coupled under the Weyl rescaling $\tilde G_{MN}=\Omega^2 G_{MN}$ with $\Omega=\left(M_5^3 / 2 F\right)^{1 / 3}$. 

In five dimensions the curvature and the extrinsic curvature of the branes transform as
\begin{equation}
\begin{aligned}
\tilde R&=\Omega^{-2}\big[R-8\,\Box\ln\Omega-12\,(\nabla\ln\Omega)^2\big],\\
\tilde K&=\Omega^{-1}\big[K+4\,n\cdot\nabla\ln\Omega\big],
\end{aligned}
\label{eq:weyl5D}
\end{equation}
with $n$ the unit normal. The $\Box\ln\Omega$ term is a total derivative whose boundary piece, $-4M_5^3\int\sqrt{-g_{\rm ind}}\,n\cdot\nabla\ln\Omega$, is cancelled exactly by the $+4M_5^3\int\sqrt{-g_{\rm ind}}\,n\cdot\nabla\ln\Omega$ generated by the GHY term, leaving the standard Einstein-frame GHY term. The potential and tensions rescale as $V=(M_5^3/2F)^{5/3}\tilde V$, $\lambda_b=(M_5^3/2F)^{4/3}\tilde\lambda_b$, so the action is again that of a minimally coupled scalar, Eq.~\eqref{eq:fullaction}.

On the background $\Omega=\Omega(\tilde\Phi(y))$, and restoring axial gauge with $dz=dy/\Omega$ absorbs the difference into $A_E$
\begin{equation}
    A_E=A_J+\ln \Omega,
    \label{eq:AE}
\end{equation}
where $A_E$ and $A_J$ are the warp factors in the Einstein and Jordan frames, respectively. Therefore, the gravitational action in Eq.~\eqref{eq:genaction} is general and applies to the case in Ref.~\cite{Gonzalo:2022jac}, as well as to many models throughout the literature. See App.~\ref{app:action} and App.~\ref{app:weyl} for more details.

To determine the amplitude at tree level, we then expand Eq.~\eqref{eq:genaction} in modes to cubic order, which gives 

\begin{widetext}
\begin{equation}
\begin{aligned}
S_{\mathrm{grav}}^{(3)}=4\sum_{n, m, l} \frac{\chi_{n m l}}{\left(\Lambda_n \Lambda_m \Lambda_l\right)^{1 / 3}}  \int d^4x
\bigg[&\frac{1}{4}\, h_{(n)}^{\mu\nu}\, \partial_\mu h_{(m)}^{\rho\sigma}\,
\partial_\nu h_{\rho\sigma}^{(l)}
- h_{(n)}^{\mu\nu}\, \partial_\nu h_{\rho\sigma}^{(m)}\,
\partial^\sigma h_{(l)\mu}^{\rho}
-\frac{1}{2}\, h_{(n)}^{\mu\nu}\, \partial_\rho h_{\nu\sigma}^{(m)}\,
\partial^\sigma h_{(l)\mu}^{\rho}
\\[2pt]
&+\frac{1}{2}\, h_{(n)}^{\mu\nu}\, \partial_\sigma h_{\nu\rho}^{(m)}\,
\partial^\sigma h_{(l)\mu}^{\rho}
+\frac{1}{12}\left(m_n^2+m_m^2+m_l^2\right) h_{(n)}^{\mu\nu}\,
h_{(m)\nu}^{\rho}\, h_{\mu\rho}^{(l)}\bigg].
\end{aligned}
\label{eq:cubic4D}
\end{equation}
\end{widetext}
Here $\chi_{nml}$ is the totally symmetric overlap between the graviton mode functions $\psi_n(y)$. $\Lambda_n$ is the coupling scale.  

The only term in which derivatives are contracted together -- which would give $M=m_n$ in Eq.~\eqref{eq:channelrate} -- is the fourth term. On external legs it gives $\mathrm{tr}\left(h^{(n)} h^{(m)} h^{(l)}\right)$ times $p_m \cdot p_l$. Since $\chi_{nml}$ and the trace are symmetric, only the symmetrized coefficient contributes, which gives
\begin{equation}
\tfrac13\big(p_n\cdot p_m+p_m\cdot p_l+p_l\cdot p_n\big)
=\tfrac16\big(m_n^2+m_m^2+m_l^2\big)\,,
\label{eq:momproduct}
\end{equation}
where we used momentum conservation on-shell. Including the prefactor $1/2$, the two derivatives give $(i)^2=-1$, so the fourth term is $-\frac{1}{12}\left(m_n^2+m_m^2+m_l^2\right) \operatorname{tr}\left(h^3\right)$ on shell. It cancels the mass term exactly. The three surviving terms have derivatives contracted with polarizations. 

The gravitational action leading to Eq.~\eqref{eq:cubic4D}, and the resulting cancellation, is general and can be understood as previewed in Sec.~\ref{sec:Intro} (see Eq.~\eqref{eq:eomcancel}) as the result of the massless 5D graviton. Writing $g_{\mu \nu} = \eta_{\mu \nu} + H_{\mu \nu}$, these two terms descend from the 5D action before expanding in modes. One comes from $R[g]$, while the other comes from the $\partial_y g \partial_y g$ term: 
\begin{align}
\mathcal{V}_4&=\frac{M_5^3}{2}\int d^4x\,dy\;\tfrac12\,e^{-2A}\,
H^{\mu\nu}\,\partial_\sigma H_{\nu\rho}\,\partial^\sigma H_\mu{}^\rho\,,
\\
\mathcal{V}_y&=\frac{M_5^3}{2}\int d^4x\,dy\;\tfrac12\,e^{-4A}\,
H^{\mu\rho}\,\partial_y H_{\mu\nu}\,\partial_y H_\rho{}^\nu\,.
\end{align}
Inserting the mode expansion $H_{\mu\nu}=\sum_n h^{(n)}_{\mu\nu}(x)\,\psi_n(y)$,
$\mathcal{V}_4$ gives the fourth term of \eqref{eq:cubic4D} with the overlap
$\int dy\,e^{-2A}\psi_n\psi_m\psi_l$, while $\mathcal{V}_y$ gets the overlap
$\int dy\,e^{-4A}\psi_n\psi_m'\psi_l'$. Integrating by parts and using the mode
equation, which applies for any warp factor,
\begin{equation}
    \partial_y\left(e^{-4 A} \psi_n^{\prime}\right)=-m_n^2 e^{-2 A} \psi_n,
    \label{eq:modes}
\end{equation}
gives the fifth term. Indeed, the cancellation can be seen in the 5D picture without using the mode expansion. Integrating by parts in $x^\sigma$ and in $y$, their sum is 
\begin{equation}
\begin{aligned}
-\frac{M_5^3}{8}\int d^4x\,dy\;
H^{(1)\mu\nu}\,&H^{(3)}{}_{\mu}{}^{\rho}
\Big[e^{-2A}\,\Box_4 H^{(2)}_{\nu\rho}\\
&+\partial_y\big(e^{-4A}\partial_y H^{(2)}_{\nu\rho}\big)\Big]\;=\;0.
\end{aligned}
\label{eq:5Dcancel}
\end{equation}
Due to symmetry the asymmetric pieces cancel pairwise when integrating by parts. The bracket is the linearized equation of motion, which vanishes on external legs for the massless 5D graviton. Naturally, the vanishing of the bracket is equivalent to the mode equation in Eq.~\eqref{eq:modes}, which caused the cancellation in the 4D picture.

\section{The Suppressed Cascade Rate}
\label{sec:suppressedrate}

Here we obtain universal scaling relations for KK graviton cascade rates. We calculate precisely below, but the answer can be understood roughly as the squared amplitude, times the two-body phase space $q/(4\pi m_n)$, times the number of decay channels. For a tower with many modes the number of modes below the decaying particle is roughly $m_n/\mkk$. Note that with KK violation up to $\delta n$, the number of decay modes is $m_n\,\delta n/\mkk$.

The momentum of the decay products in the parent's rest frame is: 
\begin{equation}
    q=\frac{1}{2 m_n} \sqrt{\lambda\left(m_n^2, m_m^2, m_l^2\right)} \ .
    \label{eq:daughterq}
\end{equation}
In Eq.~\eqref{eq:daughterq} we used the standard ``K\"all\'en'' factor  
\begin{equation}
    \lambda=\left[m_n^2-\left(m_m+m_l\right)^2\right]\left[m_n^2-\left(m_m-m_l\right)^2\right].
\end{equation}
Near threshold $\left(m_n-m_m-m_l\right)\equiv \delta m$, which gives
\begin{equation}
    \delta m =\frac{q^2}{2 m_m}+\frac{q^2}{2 m_l}=\frac{q^2}{2 \mu}, \quad \mu=\frac{m_m m_l}{m_m+m_l},
    \label{eq:qthreshold}
\end{equation}
so $q=\sqrt{2\mu\,\delta m}$ with $\mu$ the reduced mass of the decay products, and $\delta m = \delta n\,\mkk$. 

Derivatives contracting with one another, if they had survived, would have given an unsuppressed rate and allowed for rapid cascades as in \cite{Gonzalo:2022jac}. The first three terms in Eq.~\eqref{eq:cubic4D}, which do not cancel, take the form:
\begin{equation}
    h_{(n)}^{\mu \nu} \partial_\mu h_{(m)}^{\rho \sigma} \partial_\nu h_{\rho \sigma}^{(l)} \implies \varepsilon_n^{\mu \nu} p_{m \mu} p_{l \nu} \varepsilon_m^{\rho \sigma} \varepsilon_{l \rho \sigma} ,
    \label{eq:struct1}
\end{equation}
\begin{equation}
    h_{(n)}^{\mu \nu} \partial_\nu h_{\rho \sigma}^{(m)} \partial^\sigma h^{(l) \rho}{ }_\mu \implies p_m^\nu \varepsilon_{\nu \mu}^{(n)} \varepsilon^{(l) \mu \rho} \varepsilon_{\rho \sigma}^{(m)} p_l^\sigma ,
\end{equation}
\begin{equation}
    h_{(n)}^{\mu \nu} \partial_\rho h_{\nu \sigma}^{(m)} \partial^\sigma h^{(l) \rho}{ }_\mu \implies \varepsilon^{(n) \mu \nu} \varepsilon_{\nu \sigma}^{(m)} p_l^\sigma p_m^\rho \varepsilon_{\rho \mu}^{(l)} .
\end{equation}
As seen in Eq.~\eqref{eq:5Dcancel}, the fourth term, where derivatives contract, cancels with the fifth term
in Eq.~\eqref{eq:cubic4D}:
\begin{equation}
    h_{(n)}^{\mu \nu} \partial_\sigma h_{\nu \rho}^{(m)} \partial^\sigma h^{(l) \rho}{ }_\mu \implies \left(p_m \cdot p_l\right) \varepsilon^{(n) \mu \nu} \varepsilon_{\nu \rho}^{(m)} \varepsilon^{(l) \rho}{ }_\mu .
\end{equation}
The massless graviton eliminated the leading term.
We now derive scaling laws for these surviving terms. In the rest frame of the parent, the momenta are $p_n = (m, \vec 0)$ and $p_{m, l} = (E_{m, l}, \pm \vec q)$. Transverse polarizations satisfy $\epsilon_{\pm}^{(i)} \cdot p_j = 0$. The longitudinal polarization, $\epsilon_L^{(i)}=(q/m_i,0,0,E_i/m_i)$, when contracted with a different leg's momentum, is enhanced for light decay products. The results of this,  calculated in  App.~\ref{app:polarizations}, are written below:
\begin{center}
\begin{tabular*}{\columnwidth}{@{\extracolsep{\fill}}lccc@{}}
\hline
 & result & $m_m\simeq m_l$ & $m_l\ll m_m$ \\
\hline
$\epsilon_\pm^{(i)}\!\cdot p_j$ & $0$ & $0$ & $0$ \\
$\epsilon_L^{(n)}\!\cdot p_{m,l}$ & $\pm q$ & $\pm q$ & $\pm q$ \\
$\epsilon_L^{(m)}\!\cdot p_{n,l}$ & $-q\,\tfrac{m_n}{m_m}$ & $-2q$ & $\simeq -q$ \\
$\epsilon_L^{(l)}\!\cdot p_{n,m}$ & $-q\,\tfrac{m_n}{m_l}$ & $-2q$ & $-q\,\tfrac{m_n}{m_l}$ \\
$\epsilon_L^{(m)}\!\cdot\epsilon_L^{(l)}$ & $-\tfrac{q^2+E_mE_l}{m_mm_l}$ & $\simeq-1$ & $\simeq-1$ \\
\hline
\end{tabular*}
\end{center}
The surviving tensor structures, for comparable mass decay products, give something of order $q^2$ instead of order $m_n^2$. Although the rate is largest when a light decay product is emitted, we will see that comparable mass decay products dominate the cascade rate down the tower. 

Taking $m_l$ to be the lighter mode, we have $m_l \le m_m$. The largest term scales as
\begin{equation}
    \mathcal{M} \sim \frac{\chi_{nml}}{\Lambda}\left(\frac{m_n}{m_l}\right)^{2} q^2 ,
    \label{eq:Mgeneral}
\end{equation}
where $\Lambda$ is the coupling scale of the KK modes, Eq.~\eqref{eq:Lambdadef}. In Ref.~\cite{Gonzalo:2022jac} it is $\Mpl$. The $\chi$ term, depending on the wavefunction overlap, characterizes the intrinsic KK violation in the theory. At large $n$, for a smooth bulk background and for $\delta n$ small compared to $l$,  the associated KK violation is $|\chi_{nml}|\propto(\delta m)^{-2}$ with $\delta m=m_n-m_m-m_l \simeq \delta n\,\mkk$, independent of $n$ (App.~\ref{app:overlaps}). For example, for RS at strong and weak warping, respectively \cite{Lee:toappear}, 
\begin{equation}
    \left|\chi_{n m l}\right| \approx \frac{3}{4 \pi^2} \frac{1}{\left(\delta n-\frac{1}{4}\right)^2}, \qquad |\chi_{n m l}| \approx \frac{3}{2 \pi} \frac{k r_c}{\delta n^2},
    \label{eq:chilimits}
\end{equation}
where in the strongly warped case the masses are zeros of $J_1$, so $\delta m\simeq(\delta n-\tfrac14)\,\mkk$. In such cases $\chi_{nml}$ determines the KK violation \footnote{This applies when KK number is proportional to $n$, and more generally with $\delta n$ read as the change in KK number.}, and we expect this holds generally. For simplicity we take $\delta m\sim\delta n\,\mkk$ below, which does not change the result. Inserting $q^2=2\mu\,\delta m$, we find 
\begin{equation}
    \Gamma_{n\to m,l}=\frac{q\,|\mathcal{M}|^2}{8\pi m_n^2}
    \sim \frac{1}{(\delta n)^{3/2}}\,\frac{m_n^2\,\mkk^{5/2}}{\Lambda^2}\,\frac{\mu^{5/2}}{m_l^{4}} .
    \label{eq:Gammanml}
\end{equation}
Taking 
 $\tfrac{1}{2}m_l \le \mu \le m_l$, so $\mu \sim m_l$, which gives
\begin{equation}
    \Gamma_{n \rightarrow m, l} \sim \frac{1}{(\delta n)^{3 / 2} l^{3 / 2}} \frac{m_n^2 m_{\mathrm{KK}}}{\Lambda^2} .
\end{equation}
The total decay rate is then determined by summing over $\delta n$ and $l$, which yields
\begin{equation}\label{eq:Gamma_n}
\begin{aligned}
    \Gamma_n&=\sum_{l, \delta n} \Gamma_{n \rightarrow m, l} \sim \zeta\left(\tfrac{3}{2}\right)\tfrac{m_n^2 m_{\mathrm{KK}}}{\Lambda^2} \sum_{l=1}^{n / 2} \frac{1}{l^{3 / 2}} \\
    &= \zeta\left(\tfrac{3}{2}\right)\left[\zeta\left(\tfrac{3}{2}\right)-\zeta\left(\tfrac{3}{2}, \tfrac{n}{2}+1\right)\right]\tfrac{m_n^2 m_{\mathrm{KK}}}{\Lambda^2}\sim \tfrac{m_n^2 m_{\mathrm{KK}}}{\Lambda^2},
\end{aligned}
\end{equation}
where $\zeta(a)$ and $\zeta(a,b)$ are the Riemann and Hurwitz zeta functions, respectively. The \textit{decay} rate favors $\delta n \sim 1$ because $\zeta(\tfrac{3}{2})$ is $O(1)$, hence dominated by the first term. Extreme mass ratio decay products are favored because the sum over $l$ is dominated by small $m_l$, hence large $m_m$. In other words, the coefficient is $O(1)$, with no $n$ dependence, because the sum is dominated by small $l$.

In stark contrast, the \textit{cascade} proceeds through decay products of comparable mass. To see this, note that
\begin{equation}
\begin{aligned}
    \frac{d m_n}{d t}&\simeq -\sum_{l, \delta n} m_l \Gamma_{n \rightarrow m, l} \sim - \zeta(\tfrac{3}{2})\tfrac{m_n^2 m_{\mathrm{KK}}^2}{\Lambda^2} \sum_{l=1}^{n / 2} \frac{1}{l^{1 / 2}} \\
    &\simeq -\zeta(\tfrac{3}{2}) \left[ \zeta(\tfrac{1}{2}) - \zeta(\tfrac{1}{2},\tfrac{n}{2}+1)\right]\tfrac{m_n^2 m_{\rm{KK}}^2}{\Lambda^2}\\
    &\sim -\tfrac{m_n^{5/2}m_{\rm{KK}}^{3/2}}{\Lambda^2},
\end{aligned}
\end{equation}
where we used the fact that at large $n$, the Hurwitz zeta function satisfies $\zeta(s,n)\sim \frac{n^{1-s}}{s-1}$. For the same reason as before, the cascade is dominated by $\delta n \sim 1$. However -- in contrast to the decay rate -- the sum over $l$ is dominated by large $l$, with $\delta n \sim 1$, leading to an additional factor of  $O(\sqrt{n})$. Comparable mass ratio decay products dominate the cascade. The cascade is dominated by these large $l$ and $m$, giving the additional  $n$-dependent coefficient.

A universal cascade rate then follows immediately: 
\begin{equation}
\boxed{\Gamma^{\mathrm{tot}}_{\mathrm{casc}} \sim  \frac{m_n^{3 / 2} m_{\mathrm{KK}}^{3 / 2}}{\Lambda^2} \implies m_n(t) \propto t^{-2/3}.}
    \label{eq:corrected}
\end{equation}
This \textit{cascade} rate, not to be confused with the total decay rate of the $n$-th mode summed over all channels as in Eq.~\eqref{eq:Gamma_n}, predicts the movement of mass down the tower. This universal result disagrees with $m(t) \propto t^{-2/7}$ in \cite{Gonzalo:2022jac}. 

We also obtain a universal estimate for the loss of mass to kinetic energy as modes cascade down the tower. Each decay releases $E_{n\to m,l}=\delta m\simeq\delta n\,\mkk$, and by identical reasoning the rate of mass loss in the tower is
\begin{equation}
    \dot E_n=\sum_{l,\delta n}E_{n\to m,l}\,\Gamma_{n\to m,l}\sim n^2\ln n\;\frac{\mkk^4}{\Lambda^2} .
    \label{eq:Edot}
\end{equation}
This estimate assumes $\delta n\lesssim l$, because $\chi_{nml}$ is suppressed more than $(\delta n)^{-2}$ when $\delta n>l$. Due to this, Eq.~\eqref{eq:Edot} is slightly less robust than the other scaling relations.

We now compare our decay rate to the literature.
For decay products of comparable mass, $q\sim\sqrt{m_n\,\delta m}$, hence
\begin{equation}
    \frac{q}{m_n} \sim \frac{\sqrt{m_n \delta m}}{m_n}=\left(\frac{\delta m}{m_n}\right)^{1 / 2} \sim\left(\frac{\delta n \,m_{\mathrm{KK}}}{m_n}\right)^{1 / 2}.
\end{equation}
\noindent The leading term satisfies Eq.~\eqref{eq:Mgeneral}, so at \textit{most}
\begin{equation}\label{eq:leadingterm}
    \Gamma_{n \rightarrow m, l}=\frac{q|\mathcal{M}|^2}{8 \pi m_n^2} \sim \chi^2 \frac{m_n^3}{\Lambda^2}\left(\frac{\delta n\, m_{\mathrm{KK}}}{m_n}\right)^{5/2} .
\end{equation}
Consistently, multiplying Eq.~\eqref{eq:leadingterm} by $n$  channels with $\delta n \sim 1$, with $\chi \sim (\delta n)^{-2}$, while assuming comparable mass decay products gives the same rate as Eq.~\eqref{eq:corrected}. 

This Eq.~\eqref{eq:leadingterm} agrees with the equal mass decay rate of de Giorgi and Vogl~\cite{deGiorgi:2021xvm} in RS, a special case of our result Eq.~\eqref{eq:vogl}. Near threshold, this gives:
\begin{equation}
\Gamma_{n \to m,m} 
\;\simeq\;\chi^2\frac{31}{2\pi} \frac{m_n^3}{\Lambda^2} \left( \frac{q}{m_n}\right)^5 \sim \chi^2 \frac{m_n^3}{\Lambda^2}\left(\frac{\delta n m_{\mathrm{KK}}}{m_n}\right)^{5 / 2},
\label{eq:voglthresh}
\end{equation}
where $\Lambda$ is the warped-down coupling scale. Our general result helps to resolve an active tension in the literature between Giudice et al.~\cite{Giudice:2017fmj} and de Giorgi and Vogl~\cite{deGiorgi:2021xvm}. 

We find that their result \cite{deGiorgi:2021xvm}, which they obtained for the RS model, holds universally (Eq.~\eqref{eq:generalrate}): 
\begin{equation}
\begin{aligned}
\Gamma_{n\to m,l}
&=\frac{\chi^2_{nml}}{\Lambda^2}\,\frac{q^{5}\,\mathcal{P}(m_n,m_m,m_l)}{540\,\pi\,m_n^{2}\,m_m^{4}\,m_l^{4}},\\
\mathcal{P} &=\sum_i m_i^{8}+26\sum_{i\neq j}m_i^{6}m_j^{2}+126\sum_{i<j}m_i^{4}m_j^{4}\\
&\quad+364\sum_{i}m_i^{4}m_j^{2}m_k^{2}.
\end{aligned}
\label{eq:generalrate}
\end{equation}
This rate features the same $q^5$ scaling as Eq.~\eqref{eq:voglthresh}. Here one factor of $q$ arises from the phase space, while the other comes from the correction factor of $(q/m_n)^4$.

We now compare to the result of \cite{Gonzalo:2022jac}.
Summing over channels up to $\delta n_{\max}$, \cite{Gonzalo:2022jac} obtains
\begin{equation}\label{eq:wrongdecayrate}
    \Gamma^{\mathrm{tot}}_{\rm{GMOV}} \sim \frac{\beta^2\left(\delta n_{\max }\right)^{3 / 2} m_n^{7 / 2}}{M_{\mathrm{Pl}}^2 m_{\mathrm{KK}}^{1 / 2}},
\end{equation}
which gives their $m(t) \propto t^{-2/7}$ from $\Gamma^{\rm tot}t \sim 1$. 

Their findings clearly disagree with our result in Eq.~\eqref{eq:corrected}.
In Eq.~\eqref{eq:wrongdecayrate} derivatives in the three-point vertex are incorrectly assumed to contract with each other (instead of with polarizations) so that $\mathcal{M} \sim \beta m_n^2 / M_{\mathrm{Pl}}$. These terms cancel in the final rate, due to the massless graviton, so their result overestimates the decay rate of each channel by a large factor 
\begin{equation}
    \frac{\Gamma_{n \rightarrow m, l}}{\Gamma_{n \rightarrow m, l}^{\mathrm{GMOV}}} \sim \frac{\chi^2}{\beta^2} \frac{M_{\mathrm{Pl}}^2}{\Lambda^2}\left(\frac{\delta n m_{\mathrm{KK}}}{m_n}\right)^2 \simeq \left(\frac{q}{m_n}\right)^4.
\end{equation}
Note that the physical comparison is a ratio of their decay rate  to our cascade rate of Eq.~\eqref{eq:corrected}. Dividing Eq.~\eqref{eq:corrected} by Eq.~\eqref{eq:wrongdecayrate} gives 
\begin{equation}
\frac{\Gamma_{\mathrm{casc}}^{\mathrm{tot}}}{\Gamma^{\mathrm{tot}}_{\rm{GMOV}}} \sim \frac{1}{\beta^2\left(\delta n_{\max }\right)^{3 / 2}}\left(\frac{m_{\mathrm{KK}}}{m_n}\right)^2 \sim 10^{-24},
\end{equation}
so the true cascade lifetime for their assumed GeV modes with  their $m_{\rm KK}\sim \rm{eV}$, $\delta n_{\rm max} \sim  10^2$, and $\chi_{nml}\to \beta \sim 20$, would be $\tau \sim 10^{26}$ s instead of $\tau_0\sim10^{2\text{--}3}$ s.  The suppression $\left(q / m_n\right)^4 \sim\left(\delta n \, m_{\mathrm{KK}} / m_n\right)^2$, is $10^{-18}$ for $\delta n \sim 1$, $m_{\rm{KK}} \sim \rm eV$, and $m_n \sim \rm GeV$. The factors $\beta$ and $\delta n$ were needed in \cite{Gonzalo:2022jac} to reduce an overly decay rate to the standard model, even without our additional kinematical suppression, as a rapid cascade helps to reach the MeV scale before recombination to evade \cite{Slatyer:2016qyl}. One could attempt to rescue the model by considering vastly different parameters that give an initial KK mass much less than a GeV, although it must then confront other experimental constraints, such as KK-mode-mediated fifth forces and the dark matter kick velocity from the cascade.

Higher-derivative operators would not be enough to enhance the cascade decay rate, see App.~\ref{app:polarizations}.  In a companion paper we consider the question of whether CGDM could be made consistent with observations in a warped five-dimensional alternative.

\section{Discussion}\label{sec:discussion}

In this letter we have resolved the form of the  decay rate of KK graviton modes to two other KK gravitons and demonstrated a universal suppression at threshold, validating the result in   \cite{deGiorgi:2021xvm} which, however,  applied only to  the particular case of RS models. Here we have argued for the universality of the result by absorbing non-minimally coupled scalar fields into the warp-factor, yielding  Eq.~\eqref{eq:generalrate} and demonstrating the cancellation responsible for suppressing the rate.

We furthermore obtained scaling relations for these cascades, modulo the assumptions we delineated in App.~\ref{app:overlaps}. Our relations characterize the cascade down the tower (Eq.~\eqref{eq:corrected}), the decay rate of individual channels (Eq.~\eqref{eq:Gammanml}), the total decay rate of an individual mode (Eq.~\eqref{eq:Gamma_n}), and the mass loss rate in the tower (Eq.~\eqref{eq:Edot}). This gives $m(t) \propto t^{-2/3}$, in contrast to $m(t)\propto t^{-2/7}$ in  \cite{Gonzalo:2022jac,Vafa:2024fpx}. Note that the particular form of the KK violation does not affect the main result, which is  the $(q/m_n)^4$ suppression in graviton decays to two lighter gravitons, and would apply to any  Cascading Graviton Dark Matter (CGDM) model.

Although ruling out the model of \cite{Gonzalo:2022jac},  the question of whether any cascading KK modes could yield a successful dark matter scenario remains open. In Ref. \cite{Lee:toappear}, we consider an alternative RS scenario. We find that  warping can rescue some parameter space, but even so such models are  still severely constrained. Other possibilities include KK towers of scalars, but due to the required relations among couplings, they might be even less well motivated.

\begin{acknowledgments}
We thank Akshay Ghalsasi, Andreas Karch, Kevin Langhoff, Rashmish Mishra, Sonia Paban, Matt Reece, and Matt Strassler for helpful comments and conversations. VL is supported by the Network for Neutrinos, Nuclear Astrophysics and Symmetries (N3AS) through the National Science Foundation Physics Frontier Center, Grant No. PHY-2020275. LR and MR were supported by the \textbf{Gra}vity, \textbf{S}pacetime, and \textbf{P}article Physics (GRASP) Initiative at Harvard University.
\end{acknowledgments}

\appendix
\section{Gravitational Action for General Warp Factor}\label{app:action}

We consider the metric ansatz
\begin{equation}
ds^2=G_{MN}dx^Mdx^N=\hat g_{\mu\nu}(x,y)\,dx^\mu dx^\nu+dy^2\,,
\label{eq:ansatz}
\end{equation}
with $\hat g_{\mu\nu}=e^{-2A(y)}g_{\mu\nu}(x,y)$. Hatted quantities are
built from the induced metric:
$\hat K_{\mu\nu}\equiv\tfrac12\partial_y\hat g_{\mu\nu}$,
$R[\hat g]=e^{2A}R[g]$, $\sqrt{-\hat g}=e^{-4A}\sqrt{-g}$. The extrinsic
curvature reads
\begin{equation}
\hat K^\mu{}_\nu=-A'(y)\,\delta^\mu{}_\nu+K^\mu{}_\nu\,,
\qquad
K^\mu{}_\nu\equiv\tfrac12\,g^{\mu\rho}\partial_y g_{\rho\nu}\,.
\label{eq:Ksplit}
\end{equation}
The branes satisfy the Israel junction conditions (IJC), with stress
tensor $T_{ab}={\rm diag}(-\sigma,p,p,p)$:
\begin{equation}
\big[\hat K_{\mu\nu}\big]
=-\kappa_5^2\left(T_{\mu\nu}-\tfrac13\,\hat g_{\mu\nu}\,T\right),
\qquad
\kappa_5^2\equiv M_5^{-3}\,,
\label{eq:IJC}
\end{equation}
which for pure-tension branes ($p=-\sigma$,
$T^{(k)}_{ab}=-\sigma_k\hat g_{ab}$) gives
$[A']_k=\kappa_5^2\sigma_k/3$. The full action is
\begin{equation}
\begin{aligned}
S=\frac{M_5^3}{2}&\int d^4x\int dy\,\sqrt{-G}\,R_5
+S_0+S_\pi + S_{\mathrm{GHY}}
\\
&-\int d^4x\int dy\,\sqrt{-G}\Big[M_5^3\Lambda_5
+\tfrac12 G^{MN}\partial_M\Phi\,\partial_N\Phi
\\
&\qquad\qquad+V(\Phi)\Big] .
\end{aligned}
\label{eq:fullaction}
\end{equation}
The pure-tension brane actions are
$S_k= -\sigma_k\int_{y_k}d^4x\sqrt{-\hat g}$. These brane actions, as
well as the Gibbons--Hawking--York term $S_{\mathrm{GHY}}$, are required
to make Einstein's equations self-consistent. For a circle, these terms
are absent. In Appendix~\ref{app:weyl} we show this action is the most
general two-derivative starting point.

The Gauss--Codazzi identity for the ansatz \eqref{eq:ansatz} is
\begin{equation}
\begin{aligned}
\sqrt{-G}\,R_5=\sqrt{-\hat g}\Big(R[\hat g]&+\hat K^2
-\hat K^\mu{}_\nu\hat K^\nu{}_\mu\Big)\\
&-2\,\partial_y\big(\sqrt{-\hat g}\,\hat K\big)\,.
\end{aligned}
\label{eq:GC}
\end{equation}
If there is a boundary, the total-derivative term cancels exactly with
$S_{\mathrm{GHY}}$. If there is no boundary, it vanishes because it is a
total derivative. Note that
\begin{equation}
\begin{aligned}
    \hat{K}^2-\hat{K}^\mu{ }_\nu \hat{K}^\nu{ }_\mu=\left(K^2-K^\mu{ }_\nu K^\nu{ }_\mu\right)&\\
    +12 A^{\prime 2}-6 A^{\prime} K& .
\end{aligned}
    \label{eq:preKeqn}
\end{equation}
Standard identities give, from Eq.~\eqref{eq:Ksplit},
\begin{equation}
    K \equiv K^\mu{ }_\mu=\partial_y \ln \sqrt{-g}, \quad  \partial_y \ln \sqrt{-\hat{g}}=-4 A^{\prime}+K .
\end{equation}
Solving for $K$ and inserting into Eq.~\eqref{eq:preKeqn} gives
\begin{equation}
\begin{aligned}
\hat K^2-\hat K^\mu{}_\nu\hat K^\nu{}_\mu
=\big(K^2&-K^\mu{}_\nu K^\nu{}_\mu\big)\\
&-\frac{6\,\partial_y\big(A'\sqrt{-\hat g}\big)}{\sqrt{-\hat g}}
+6A''-12A'^2\,.
\end{aligned}
\label{eq:KhatA}
\end{equation}
If the boundary is absent, the total derivative vanishes. If there is a
boundary, it cancels with the brane actions because
$3 M_5^3\left[A^{\prime}\right]_k=\sigma_k$ by the IJC. If there are
distributional sources in $A''$ within the bulk, i.e.\ ``interior
branes,'' their contributions to the action cancel for the same reason.

It is quite generally true that the remaining non-distributional pieces
cancel with the matter action. Maximally symmetric slices, such as one
that models our universe, satisfy
\begin{equation}
    T^M{ }_N=\operatorname{diag}\left(p, p, p, p, p_y\right), \quad T_{\mu \nu}=p(y)\, \hat{g}_{\mu \nu} ,
\end{equation}
where $p=p(y)$ and $p_y=p_y(y)$. The Einstein tensor reads
\begin{equation}
    G_{\mu \nu}=\left(6 A^{\prime 2}-3 A^{\prime \prime}\right) \hat{g}_{\mu \nu}, \quad G_{55}=6 A^{\prime 2} .
\end{equation}
The $(\mu \nu)$ Einstein equations are
$G_{\mu \nu}=\kappa_5^2 T_{\mu \nu}$, so we immediately find
\begin{equation}
    \mathcal{L}_M=p(y)=-\frac{M_5^3}{2}\left(6 A^{\prime \prime}-12 A^{\prime 2}\right), \quad p_y=6 M_5^3 A^{\prime 2} ,
\end{equation}
as quoted in Eq.~\eqref{eq:genmatteraction}. This cancels exactly with the
remaining terms in Eq.~\eqref{eq:KhatA}, leaving only
$\left(K^2-K^\mu{ }_\nu K^\nu{ }_\mu\right)$. The same argument holds in
general dimensions, with the only surviving piece being the cosmological
constant on the brane. 

Note that the anisotropy in the pressure is
proportional to $A''$, which breaks KK number conservation
(Appendix~\ref{app:isometry}):
\begin{equation}
    p_y-p=3 M_5^3 A^{\prime \prime} .
\end{equation}
More generally this reads
\begin{equation}
    p_y-p=(d-1) M_*^{d-1}\left[A^{\prime \prime}-\frac{{R[\hat{g}]}}{d(d-1)} \right] ,
    \label{eq:anisotropy}
\end{equation}
with $M_*$ the higher-dimensional Planck mass ($M_*=M_5$ for $d=4$) and $R[\hat{g}]$ the curvature of the slice.

Let us see how this cancellation occurs in a specific example. The
background equations with a $\Phi$ field for $\Lambda_4=0$
are~\cite{DeWolfe:1999cp}
\begin{equation}
6A'^2=-\Lambda_5+\kappa_5^2\Big(\tfrac12\bar\Phi'^2-V\Big)\,,
\qquad
3A''=\kappa_5^2\,\bar\Phi'^2\,.
\label{eq:DFGK}
\end{equation}
For a minimally coupled scalar, with
$\Phi=\bar{\Phi}(y)+\varphi(x, y)$, the matter Lagrangian on the
background reads
\begin{equation}
\begin{aligned}
\mathcal{L}_M
&=-M_5^3\Lambda_5-\tfrac12\bar\Phi'^2-V(\bar\Phi)
-\sum_k\sigma_k\,\delta(y-y_k)\\
&=-\frac{M_5^3}{2}\big(6A''-12A'^2\big)\,,
\end{aligned}
\label{eq:mattercancel}
\end{equation}
by \eqref{eq:DFGK}, where $\sum_k \sigma_k \delta\left(y-y_k\right)$ are
the brane tension terms. This cancels exactly against the $A''$ and
$A'^2$ terms of \eqref{eq:KhatA}, a pattern which holds much more
generally, as shown above.

Consider the surviving terms. From Eq.~\eqref{eq:Ksplit} we have
$K^2-K^\mu{}_\nu K^\nu{}_\mu
=\tfrac14\big(g^{\mu\nu}g^{\rho\sigma}-g^{\mu\rho}g^{\nu\sigma}\big)
\partial_yg_{\mu\nu}\,\partial_yg_{\rho\sigma}$ and
$\sqrt{-\hat g}=e^{-4A}\sqrt{-g}$, $R[\hat g]=e^{2A}R[g]$, which
reproduces Eq.~\eqref{eq:genaction}. The zero mode
($\partial_yg_{\mu\nu}=0$) gives
\begin{equation}
\begin{aligned}
M_{\rm Pl}^2&=M_5^3\int dy\,e^{-2A(y)}\\
&\xrightarrow{\,A=k|y|\,}\;
\frac{M_5^3}{2k}\left(1-e^{-2\pi kr_c}\right),
\end{aligned}
\label{eq:planck}
\end{equation}
specializing to RS in the last step. In fact, this holds not only for general $A(y)$
but also for general couplings after a Weyl rescaling, which we now
show.

\section{General Couplings Reduce to Warp Factors}\label{app:weyl}

The most general two-derivative action for the scalar is
\begin{equation}
\begin{aligned}
S=\int d^4x\,dy\,&\sqrt{-\tilde G}\Big[F(\tilde\Phi)\,\tilde R\\
&-\tfrac12\,k(\tilde\Phi)\,\tilde G^{MN}\partial_M\tilde\Phi\,\partial_N\tilde\Phi
-\tilde V(\tilde\Phi)\Big]\\
&-\sum_b\int d^4x\,\sqrt{-\tilde g_{\rm ind}}\;\tilde\lambda_b(\tilde\Phi)
+\tilde S_{\rm GHY}\,,
\end{aligned}
\label{eq:jordan}
\end{equation}
with
$\tilde S_{\rm GHY}=2\int\sqrt{-\tilde g_{\rm ind}}\;F(\tilde\Phi)\,\tilde K$.
Weyl transform $\tilde G_{MN}=\Omega^2(\tilde\Phi)\,G_{MN}$. In $D$
dimensions the standard identities hold:
\begin{align}
\tilde R&=\Omega^{-2}\Big[R-2(D-1)\,\Box\ln\Omega
\nonumber\\
&\qquad\qquad -(D-1)(D-2)\,(\nabla\ln\Omega)^2\Big],
\\
\tilde K&=\Omega^{-1}\Big[K+(D-1)\,(n\cdot\nabla)\ln\Omega\Big],
\label{eq:wald}
\end{align}
with $\sqrt{-\tilde G}=\Omega^D\sqrt{-G}$ and
$\sqrt{-\tilde g_{\rm ind}}=\Omega^{D-1}\sqrt{-g_{\rm ind}}$. The first
identity can be found in Wald~\cite{Wald:1984rg}, Appendix D. The second is obtained
straightforwardly using the same method.

Choose the conformal factor to scale out the prefactor of $\tilde R$,
$\Omega^{D-2} F=\frac{M_5^3}{2}$:
\begin{equation}
\Omega=\left(\frac{M_5^3}{2F}\right)^{1/3},
\quad
\nabla_M\ln\Omega=-\frac{F'}{3F}\,\partial_M\tilde\Phi\,.
\label{eq:omegachoice}
\end{equation}
For $D=5$, using Eqs.~\eqref{eq:wald} and \eqref{eq:omegachoice}, the EH
and GHY terms transform as
\begin{equation}
\begin{aligned}
\sqrt{-\tilde G}\,F\,\tilde R
=\sqrt{-G}\;\frac{M_5^3}{2}\Big[&R-8\,\Box\ln\Omega\\
&-12\,(\nabla\ln\Omega)^2\Big],
\end{aligned}
\end{equation}
\begin{equation}
    2 \sqrt{-\tilde{g}_{\text {ind }}} F \widetilde{K}=\sqrt{-g_{\text {ind }}}\, M_5^3\left[K+4\, n \cdot \nabla \ln \Omega\right].
    \label{eq:transformedGHY}
\end{equation}
The $\Box\ln\Omega$ term is a total derivative. Using the divergence
theorem, the boundary piece cancels exactly against the transformed GHY
term in Eq.~\eqref{eq:transformedGHY},
\begin{equation}
\begin{aligned}
-4M_5^3&\int\sqrt{-g_{\rm ind}}\;n\cdot\nabla\ln\Omega\\
&+M_5^3\int\sqrt{-g_{\rm ind}}\,\big(K+4\,n\cdot\nabla\ln\Omega\big)\\
&\qquad=M_5^3\int\sqrt{-g_{\rm ind}}\,K\,,
\end{aligned}
\end{equation}
leaving the standard Einstein-frame GHY term. Using
Eq.~\eqref{eq:omegachoice}, the $(\nabla\ln\Omega)^2$ term combines with
the transformed kinetic term, which gives
\begin{equation}
\mathcal{L}_{\rm kin}
=-\frac12\,\frac{M_5^3}{2F}\left[k+\frac{8}{3}\frac{F'^2}{F}\right]
(\partial\tilde\Phi)^2\,.
\label{eq:kinetic}
\end{equation}
The canonically normalized field $\hat\Phi$ then satisfies
\begin{equation}
    \left(\frac{d \hat{\Phi}}{d \tilde{\Phi}}\right)^2=\frac{M_5^3}{2 F}\left[k+\frac{8}{3} \frac{F'^2}{F}\right] .
\end{equation}
The kinetic term has positive sign (no ghosts) when $F>0$ and
$k+\tfrac{8}{3}F'^2/F>0$. Finally, the rescaled potentials and tensions
are
\begin{equation}
V(\hat\Phi)=\left(\frac{M_5^3}{2F}\right)^{5/3}\tilde V\,,
\qquad
\lambda_b(\hat\Phi)=\left(\frac{M_5^3}{2F}\right)^{4/3}\tilde\lambda_b\,.
\label{eq:rescalings}
\end{equation}
The result is the minimally coupled action
\begin{equation}
\begin{aligned}
S=\int d^4x\,dy\,&\sqrt{-G}\left[\frac{M_5^3}{2}R-\frac12(\partial\hat\Phi)^2
-V(\hat\Phi)\right]\\
&-\sum_b\int d^4x\,\sqrt{-g_{\rm ind}}\;\lambda_b(\hat\Phi)\\
&+M_5^3\int\sqrt{-g_{\rm ind}}\,K\,.
\end{aligned}
\label{eq:einsteinframe}
\end{equation}
Recalling that we wrote $\hat{g}$ for the induced metric in
Appendix~\ref{app:action}, and that the brane actions are given by
$\sum_b \int d^4 x \sqrt{-g_{\text {ind }}} \lambda_b(\hat{\Phi})$, this
is the same form considered in Eq.~\eqref{eq:fullaction}.

The metric ansatz is also identical, because the non-minimal coupling
can be absorbed into the warp factor. Take the Jordan-frame ansatz
$d\tilde s^2=e^{-2A_J(y)}\eta_{\mu\nu}dx^\mu dx^\nu+dy^2$ with a scalar
profile $\tilde\Phi(y)$, so that on the background
$\Omega=\Omega\big(\tilde\Phi(y)\big)$. The Einstein metric is
$ds^2=\Omega^{-2}d\tilde s^2$, with
\begin{equation}
ds^2=\Omega^{-2}(y)\,e^{-2A_J(y)}\,\eta_{\mu\nu}dx^\mu dx^\nu
+\Omega^{-2}(y)\,dy^2\,.
\end{equation}
To restore the axial gauge choice, $G_{55}=1$, we use
$dz=dy/\Omega(y)$, which gives Eq.~\eqref{eq:ansatz} with
\begin{equation}
A_E(z)=A_J\big(y(z)\big)+\ln\Omega\big(y(z)\big)\,,
\label{eq:warpshift}
\end{equation}
as quoted in Eq.~\eqref{eq:AE}. Thus the KK violation sourced by the
vacuum expectation value for $\Phi$ can be understood in full generality
using $A(y)$, as in Appendix~\ref{app:action}. 

We conclude that the
coupling considered throughout the literature for the decay of massive
gravitons down the tower always cancels when the 5D graviton is
massless, severely suppressing the cascade rate.

\section{KK Modes, the Mode Equation, and Coupling Scales}\label{app:modes}

Here we review the KK expansion of the action \eqref{eq:genaction}. Write
$g_{\mu\nu}=\eta_{\mu\nu}+\kappa\,h_{\mu\nu}$ with
$\kappa\equiv2M_5^{-3/2}$, so that $h_{\mu\nu}$ is canonically
normalized in five dimensions. To second order in $h$,
Eq.~\eqref{eq:genaction} gives
\begin{equation}
\begin{aligned}
S^{(2)}=2\int d^4x\,dy\,\Big[&e^{-2A}\,\mathcal{L}^{(2)}_{\rm FP}[h]\\
&-\tfrac14\,e^{-4A}\big(\partial_y h_{\mu\nu}\partial_y h^{\mu\nu}
-(\partial_y h)^2\big)\Big],
\end{aligned}
\label{eq:quadratic}
\end{equation}
where $h\equiv h^\mu{}_\mu$ and $\mathcal{L}^{(2)}_{\rm FP}$ is the
standard Fierz--Pauli Lagrangian,
\begin{equation}
\begin{aligned}
\mathcal{L}^{(2)}_{\rm FP}=&-\tfrac14\,\partial_\rho h_{\mu\nu}\partial^\rho h^{\mu\nu}
+\tfrac12\,\partial_\nu h^{\mu\nu}\partial^\rho h_{\mu\rho}\\
&-\tfrac12\,\partial_\mu h\,\partial_\nu h^{\mu\nu}
+\tfrac14\,\partial_\mu h\,\partial^\mu h\,.
\end{aligned}
\end{equation}
The linearized equation of motion for transverse-traceless
fluctuations is
\begin{equation}
e^{-2A}\,\Box_4 h_{\mu\nu}+\partial_y\big(e^{-4A}\,\partial_y h_{\mu\nu}\big)=0\,,
\label{eq:lineom}
\end{equation}
the same combination that appears in the bracket of
Eq.~\eqref{eq:5Dcancel}. Expanding in modes,
\begin{equation}
h_{\mu\nu}(x,y)=\frac{1}{\sqrt{\ell}}\sum_n h^{(n)}_{\mu\nu}(x)\,\psi_n(y)\,,
\label{eq:expansion}
\end{equation}
where $\ell$ is the length of the extra dimension ($\ell= \pi r_c$ in RS), and separating
variables with $\Box_4 h^{(n)}=m_n^2\,h^{(n)}$, Eq.~\eqref{eq:lineom}
yields Eq.~\eqref{eq:modes} for Neumann BCs: $\psi_n'|_0^{\ell}=0$. This is a
Sturm--Liouville problem. We write
\begin{equation}
\frac{1}{\ell}\int dy\,e^{-2A}\,\psi_n\psi_m=\delta_{nm}\,,
\label{eq:norm}
\end{equation}
so that the 4D modes $h^{(n)}_{\mu\nu}$ are canonically normalized.
Integrating by parts and using the mode equation gives
\begin{equation}
\frac{1}{\ell}\int dy\,e^{-4A}\,\psi_n'\psi_m'=m_n^2\,\delta_{nm}\,.
\end{equation}
For the zero mode, $\psi_0$ is
constant, and \eqref{eq:norm} reproduces the 4D Planck mass,
Eq.~\eqref{eq:planck}.

Fields localized on a brane at $y_b$ couple to the induced
metric, so to linear order in the fluctuation
\begin{equation}
S_{\rm int}=\frac{\kappa}{2}\int d^4x\;h_{\mu\nu}(x,y_b)\,T^{\mu\nu}\,,
\label{eq:braneint}
\end{equation}
with $T^{\mu\nu}$ the brane stress tensor. Inserting the expansion
\eqref{eq:expansion} and $\kappa=2M_5^{-3/2}$, the $n$-th mode couples
with strength $|\psi_n(y_b)|/(M_5^{3/2}\ell^{1/2})\equiv1/\Lambda_n$,
where
\begin{equation}
\Lambda_n\equiv\frac{M_5^{3/2}\,\ell^{1/2}}{|\psi_n(y_b)|}\,.
\label{eq:Lambdadef}
\end{equation}
For the zero mode, \eqref{eq:norm} gives
$\psi_0^2=\ell/\!\int dy\,e^{-2A}$, so
$\Lambda_0=\big(M_5^3\int dy\,e^{-2A}\big)^{1/2}=\Mpl$ by
Eq.~\eqref{eq:planck}. Note the massless graviton couples with strength
$1/\Mpl$.

At third order in $h$ this yields the cubic action \eqref{eq:cubic4D},
with the overlap coefficient normalized as
\begin{equation}
\chi_{nml}\equiv
\frac{\int dy\,e^{-2A}\,\psi_n\psi_m\psi_l}
{\ell\,\big[\psi_n(y_b)\,\psi_m(y_b)\,\psi_l(y_b)\big]^{1/3}}\,.
\label{eq:chidef}
\end{equation}
The brane values cancel in
Eq.~\eqref{eq:cubic4D} because,
\begin{equation}
\frac{\chi_{nml}}{(\Lambda_n\Lambda_m\Lambda_l)^{1/3}}
=\frac{\int dy\,e^{-2A}\,\psi_n\psi_m\psi_l}{(M_5\,\ell)^{3/2}}.
\label{eq:refindep}
\end{equation}
In RS $y_b$ is the IR brane, and all $\Lambda_n$
approach the common warped-down scale
$\Lambda\simeq\Mpl\,e^{-\pi k r_c}$, which is the scale appearing in
the consistency check of Sec.~\ref{sec:suppressedrate}. In RS $k$ is near the cutoff scale. At small $k$, KK graviton species push the EFT cutoff to the standard $\Lambda_{\mathrm{EFT}} \simeq M_5 e^{- \pi k r_c}$. 

\section{Anisotropy Breaks KK Number}\label{app:isometry}

Decays down the tower require breaking KK number, i.e.\ breaking
translations along $y$. At the two-derivative level the breaking is
measured by $A''$, which we now show is an anisotropy of the background
in the $y$ direction.

Consider
$\xi=\xi^\mu(x,y)\partial_\mu+\xi^5(x,y)\partial_y$ with
$ds^2=e^{-2A(y)}\eta_{\mu\nu}dx^\mu dx^\nu+dy^2$. The current
$J^M=T^{MN}\xi_N$ obeys
$\nabla_MJ^M=\tfrac12\,T^{MN}(\mathcal{L}_\xi G)_{MN}$. Note that KK number must be conserved when $\xi$ is an isometry. The components for the Lie derivative $\left(\mathcal{L}_{\xi} G\right)_{M N}$ are
\begin{align}
(\mathcal{L}_\xi G)_{55}&=2\,\partial_5\xi^5\,,
\nonumber\\
(\mathcal{L}_\xi G)_{\mu5}&=e^{-2A}\eta_{\mu\rho}\,\partial_5\xi^\rho
+\partial_\mu\xi^5\,,
\label{eq:liederiv}
\\
(\mathcal{L}_\xi G)_{\mu\nu}
&=e^{-2A}\Big[\partial_{(\mu}\xi_{\nu)}-2A'\xi^5\eta_{\mu\nu}\Big]\,,
\nonumber
\end{align}
with $\xi_\mu\equiv\eta_{\mu\rho}\xi^\rho$ and $\partial_{(\mu}\xi_{\nu)}\equiv\partial_\mu\xi_\nu+\partial_\nu\xi_\mu$. Setting all three to zero, we find
$\xi^5=\xi^5(x)$, and $\partial_5\xi_\mu=-e^{2A}\partial_\mu\xi^5$. Inserting these into the third condition and applying $\partial_5$:
\begin{equation}
\partial_\mu\partial_\nu\,\xi^5=-\,e^{-2A}A''\;\xi^5\,\eta_{\mu\nu}\,=-c \, \xi^5  \eta_{\mu \nu}.
\label{eq:flatcase}
\end{equation}
The left side is $y$-independent, so for $\xi^5\not\equiv0$ we
have a separation constant $e^{-2A(y)}A''(y)=c$. 

It is simple to show this vanishes for flat slices. Apply $\partial_\rho$ to \eqref{eq:flatcase}. The LHS is symmetric in $(\rho \mu \nu)$, so $c\left(\eta_{\mu \nu} \partial_\rho \xi^5-\eta_{\rho \nu} \partial_\mu \xi^5\right)=0$. Contracting $\mu \nu$ gives $c(d-1) \partial_\rho \xi^5=0$. Suppose $c \ne 0$. Then $\xi^5$ is constant, but then \eqref{eq:flatcase} gives $c=0$, a contradiction. Hence
\begin{equation}
c=0
\quad\Longrightarrow\quad
A''=0 \, ,
\label{eq:flatresult}
\end{equation}
so flat slicings have an isometry along $y$ iff $A''=0$. 

For maximally symmetric slicings the separation constant need not
vanish. The argument is simply a covariant version of the previous one, where \eqref{eq:flatcase} becomes $\nabla_\mu\nabla_\nu\xi^5=-c\,\xi^5\, g_{\mu\nu}$, and computing $\left[\nabla_\lambda, \nabla_\mu\right] \nabla_\nu \xi^5$ gives
\begin{equation}
    -c\left(g_{\mu \nu} \nabla_\lambda \xi^5-g_{\lambda \nu} \nabla_\mu \xi^5\right)=-R^\sigma{ }_{\nu \lambda \mu} \nabla_\sigma \xi^5 .
\end{equation}
Contracting with $g^{\lambda \nu}$ gives $R^\sigma{ }_\mu \nabla_\sigma \xi^5=c(d-1) \nabla_\mu \xi^5 $. For the maximally symmetric slice, $R_{\mu \nu} = (R/d)g_{\mu \nu}$, which implies $R^\sigma_\mu = (R/d) \delta^\sigma_\mu$. This yields
\begin{equation}
A''(y)=\frac{R[\hat{g}]}{d(d-1)}\,,
\label{eq:curvedresult}
\end{equation}
so AdS, dS, and flat slicings can preserve the isometry, with separation constant $c = e^{-2A}A''= \frac{2\Lambda_d}{(d-1)(d-2)}$. There is no
tension between \eqref{eq:curvedresult} and the flat-slice result: for
curved slicings the background equations become
\begin{equation}
(d-1)\left[A''-\frac{R[\hat{g}]}{d(d-1)}\right]=\kappa_5^2\,\bar\Phi'^2\,,
\label{eq:curvedDFGK}
\end{equation}
the LHS vanishes
iff $\bar\Phi'=0$. This is Eq.~(18) of Ref.~\cite{DeWolfe:1999cp}.

Therefore KK number is violated when
$A''\ne\frac{R[\hat{g}]}{d(d-1)}$. In flat spacetimes this can
arise when a brane sources a jump $[A']\neq0$ (distributional $A''$), or
when a scalar profile sources $3A''=\kappa_5^2\bar\Phi'^2$
[Eq.~\eqref{eq:DFGK}]. In each case the 5D graviton
remains massless, which is exactly the condition that underlies the
cancellation of Sec.~\ref{sec:cancel}. Note that the anisotropy in the
$y$ direction breaks KK number conservation:
\begin{equation}
\partial_\mu \partial_\nu \xi^5=-\frac{e^{-2 A}}{3 M_5^3}\left(p_y-p\right) \xi^5 \eta_{\mu \nu}\,,
\end{equation}
so
$p_y-p=\bar{\Phi}^{\prime 2}(y)+\sum_k \sigma_k \delta\left(y-y_k\right)$.

\section{Polarization Sums and the Decay Rate}\label{app:polarizations}

In the parent rest frame, $p_n = (m,\vec{0})$ and
$p_{m,l} =\left(E_{m, l}, \pm \vec{q}\right)$. The usual polarization
vectors take the form
\begin{equation}
    \epsilon_{ \pm}^\mu=\frac{1}{\sqrt{2}}(0,1, \pm i, 0), \quad \epsilon_L^\mu=(0,0,0,1) .
\end{equation}
Boosting along $z$, we have $p^\mu=\left(E_i, 0,0, q\right)$ with
$E_i=\sqrt{m_i^2+q^2}$. The longitudinal polarization is orthogonal to
the momentum:
\begin{equation}
    \epsilon_L^\mu=\left(\frac{q}{m_i}, 0,0, \frac{E_i}{m_i}\right) .
\end{equation}
The other polarizations are unaffected by the boost. The massive 4D
graviton's spin-2 polarizations are linear combinations of products of
the spin-1 polarizations:
\begin{equation}
\begin{aligned}
    \varepsilon_{\mu \nu}^{( \pm 2)}&=\epsilon_{ \pm} \epsilon_{ \pm}, \qquad \varepsilon_{\mu \nu}^{( \pm 1)}=\frac{1}{\sqrt{2}}\left(\epsilon_{ \pm} \epsilon_L+\epsilon_L \epsilon_{ \pm}\right),\\
    \varepsilon_{\mu \nu}^{(0)}&=\frac{1}{\sqrt{6}}\left(\epsilon_{+} \epsilon_{-}+\epsilon_{-} \epsilon_{+}+2 \epsilon_L \epsilon_L\right) .
\end{aligned}
\end{equation}
Possible inner products between these and the momenta are
\begin{equation}
\begin{aligned}
    \epsilon_{ \pm}^{(i)} \cdot p_j&=0 , \qquad
    \epsilon_L^{(n)} \cdot p_{m, l}=\pm q ,\\
    \epsilon_L^{(m)} \cdot p_{n,l}&=-\frac{q\left(E_m+E_l\right)}{m_m}=-q\,\frac{m_n}{m_m} ,\\
    \epsilon_L^{(m)} \cdot \epsilon_L^{(l)}&=-\frac{q^2+E_m E_l}{m_m m_l} ,
\end{aligned}
\end{equation}
with $\epsilon_L^{(l)}\cdot p_{n,m}$ obtained from the third line by $m\leftrightarrow l$. Evaluated for comparable mass decay products, $m_n\simeq2m_m\simeq2m_l$, and for a light decay product, $m_l\ll m_m\simeq m_n$, these give the entries of the table in Sec.~\ref{sec:suppressedrate}. The suppression is explicit for
$\varepsilon_n^{\mu \nu} p_{m \mu} p_{l \nu} \varepsilon_m^{\rho \sigma} \varepsilon_{l \rho \sigma}$, $\epsilon^2 \sim \mathcal{O}(1)$:
\begin{equation}
    \varepsilon_n^{(0) \mu \nu} p_{m \mu} p_{l \nu}=\frac{2}{\sqrt{6}}\left(\epsilon_L \cdot p_m\right)\left(\epsilon_L \cdot p_l\right)=-\frac{2}{\sqrt{6}} q^2 .
\end{equation}
These tensor structures always give something of order $q^2$ instead of
something of order $m_n^2$, as used in Sec.~\ref{sec:suppressedrate}.

Carrying out the polarization sums with the surviving structures gives the exact rate for general daughter masses in RS at strong warping ($\Lambda_n=\Lambda_m=\Lambda_l=\Lambda$), as in Ref.~\cite{deGiorgi:2021xvm}:
\begin{equation}
\begin{aligned}
\Gamma_{n\to m,l}
&=\frac{\chi^2_{nml}}{\Lambda^2}\,\frac{q^{5}\,\mathcal{P}(m_n,m_m,m_l)}{540\,\pi\,m_n^{2}\,m_m^{4}\,m_l^{4}},\\
\mathcal{P} &=\sum_i m_i^{8}+26\sum_{i\neq j}m_i^{6}m_j^{2}+126\sum_{i<j}m_i^{4}m_j^{4}\\
&\quad+364\sum_{i}m_i^{4}m_j^{2}m_k^{2},
\end{aligned}
\label{eq:generalrate}
\end{equation}
with $q$ the momentum of Eq.~\eqref{eq:daughterq} (the K\"all\'en factor enters as $\lambda^{5/2}=32\,m_n^5q^5$), $\mathcal{P}$ symmetric in the three masses ($i,j,k$ distinct, the last sum over the three choices of $i$), and a symmetry factor $\tfrac12$ for identical decay products. Near threshold with $m_l\ll m_n$ this gives Eq.~\eqref{eq:Gammanml}, and for $m_m=m_l$ it gives Eq.~(A.7) of Ref.~\cite{deGiorgi:2021xvm}:
\begin{equation}
\begin{aligned}
\Gamma_{n\to m,m}
=\chi^2_{nmm}&\,\frac{\big(m_n^2-4m_m^2\big)^{5/2}}
{34560\,\pi\Lambda^2 m_m^8 m_n^2}
\big[780\,m_m^6m_n^2\\
&{+}616\,m_m^4m_n^4{+}52\,m_m^2m_n^6\\
&{+}180\,m_m^8{+}m_n^8\big]\,.
\end{aligned}
\label{eq:vogl}
\end{equation}
The decay product momentum is $q=\tfrac12\sqrt{m_n^2-4m_m^2}$, so the prefactor
is $(m_n^2-4m_m^2)^{5/2}=32\,q^5$. Near threshold
($m_n\to2m_m\equiv m$) the bracket tends to $16740\,m_m^8$, yielding
Eq.~\eqref{eq:voglthresh} of the main text.

Finally, consider a curvature-squared term with an $\mathcal{O}(1)$
coefficient and the suppression scale at $M_5$. Even assuming no
cancellation at all, so that the rate remains unsuppressed,
the suppression factor $(m/M_5)^4$ gives
\begin{equation}
\begin{aligned}
\Gamma^{\rm tot}_{R^2}
&\;\lesssim\;
\left(\frac{m}{M_5}\right)^{4}\Gamma^{\rm tot}\\
&\;=\;
\beta^2\,(\delta n_{\max})^{3/2}\,
\frac{m^{15/2}}{\Mpl^{14/3}\,\mkk^{11/6}}\,,
\end{aligned}
\label{eq:hdrate}
\end{equation}
with $\Gamma^{\rm tot}$ the unsuppressed rate
\eqref{eq:wrongdecayrate}, where we used $M_5^3=\Mpl^2\,\mkk$ to
eliminate the 5D scale. For example, Gauss-Bonnet terms will be suppressed. 

\section{Overlaps at Large Mode Number}\label{app:overlaps}

In this section, we derive the dependence of the overlap $\chi_{nml}$ on $\delta n$ to find the $n$-independent $(\delta m)^{-2}$ form used in the main text. For this estimate we assume KK violation arises through a finite smooth  bulk background. With this assumption, the form works when  $m_n \gg m_{\mathrm{KK}}$, so $n$ is large. For simplicity, we also assume that $\delta m < m_l$.  Though not always true when the light state is very light, it would change the result very little when the number of light modes is small.  It is also assumed that standard Neumann boundary conditions apply at the brane, so brane localized curvature terms (e.g. DGP terms) may give model dependent corrections to this estimate.  Finally, for simplicity, we have done the analysis when branes are the only sharp features in the bulk. With additional sharp features, we would match the WKB solution outlined below to the exact solution near the feature  to determine $\theta$ on each side of it. Independently  of these assumptions, the $(q/m_n)^4$ suppression applies regardless of how $\chi_{n m l}$ scales with $\delta m$. The additional assumptions enter only into the scaling arguments.

Write $\psi_n=e^{3 A / 2} \varphi_n$ and use conformal coordinates $z(y)=\int_0^{y} e^A d y$. Let $\ell$ be the size of the extra dimension. In RS, $\ell =\pi r_c$.  Then \cite{DeWolfe:1999cp}:
\begin{equation}
    -\varphi_n^{\prime \prime}+V \varphi_n=m_n^2 \varphi_n, \quad V=\frac{9}{4} A_z^2-\frac{3}{2} A_{z z},
\end{equation}
where $V(z)$ is the usual volcano potential. 

Defining $p(z)=\sqrt{m_n^2-V(z)}$, the WKB solution to the Schrödinger equation applies at large $n$: 
\begin{equation}
    \varphi_n(z)=\frac{N}{\sqrt{p(z)}} \cos \Theta_n(z), \quad \Theta_n(z)=\int_{z_0}^z p d z^{\prime}+\alpha.
\end{equation}
At large mode number $n \gg 1$ we have $m_n^2 \gg V$ away from the UV brane, so $p \simeq m_n - V/(2 m_n)$. Normalizing using Eq.~\eqref{eq:norm}, the WKB solution reads:
\begin{equation}
\begin{aligned}
    \psi_n(y) &\simeq e^{3 A(y) / 2} \sqrt{\frac{2 \ell}{z(\ell)}} \cos \Theta_n(y), \\
    \Theta_n(y)&=m_n z(y)+\theta+O\left( \frac{1}{n}\right),
\end{aligned}
\end{equation}
where the constant $\theta$ is $n$-independent at leading order.

Inserting these modes into Eq.~\eqref{eq:chidef}, the slowly oscillating part of the phase depends on the KK violation 
\begin{equation}
    \cos \Theta_n \cos \Theta_m \cos \Theta_l=\tfrac{1}{4} \cos \left(\delta m z(y)-\theta+O\left( \tfrac{1}{n}\right)\right),
\end{equation}
where $\delta m \equiv m_n - m_m - m_l$. Then integrating by parts at large $n$, we find the simple expression
\begin{equation}
\begin{aligned}
    \chi_{nml} \simeq{}& \frac{s\, e^{-3A(\ell)/2}}{2\, z(\ell)\, \delta m^2}
    \Bigg( \Big[ \big(\partial_z e^{3A/2}\big)\cos(\delta m\, z - \theta) \Big]_{0}^{z(\ell)} \\
    &\qquad - \int_0^{z(\ell)} dz\, \big(\partial_z^2 e^{3A/2}\big)\cos(\delta m\, z - \theta) \Bigg),
\end{aligned}
\end{equation}
where $s \equiv  \pm1$ does not affect $|\chi_{nml}|$. We conclude that $|\chi_{nml}| \propto (\delta m)^{-2}$ is independent of $n$ at large $n$, which is the case of interest for cascading KK graviton models.

The reason for this scaling is that $n$ drops out of the slowly varying, approximately sinusoidal phase. Due to this, these overlaps are strongly dependent on the violation of KK number, but not on mode number.

Let us verify that this is consistent with known results. For RS, $A = k |y|$ with $\ell= \pi r_c$, so $z(\ell)=\left(e^{k \ell}-1\right) / k$ and $e^{3 A / 2}=(1+k z)^{3 / 2}$ in the bulk between the branes. Then since $A_z = e^{-A}A'(y)$ and $A_{zz} = e^{-2 A}(A''(y)-A'(y)^2)$, the strong and weak warping limits are Eq.~\eqref{eq:chilimits} of the main text, with $\delta n \equiv n - m - l$. The parameter $\theta$ is fixed at strong warping by matching the WKB solution to the exact solution near the brane, where $m_n^2<V$, and $\theta=-5\pi/4$ follows from $J_2$ asymptotics. For weak warping, $m_n^2\gg V$ there and $\theta=0$ follows from Neumann BCs. In each case $\theta$ is independent of $n$ at leading order.

\bibliographystyle{apsrev4-2}
\bibliography{biblio.bib}

@article{Gonzalo:2022jac,
    author = "Gonzalo, Eduardo and Montero, Miguel and Obied, Georges and Vafa, Cumrun",
    title = "{Dark dimension gravitons as dark matter}",
    eprint = "2209.09249",
    archivePrefix = "arXiv",
    primaryClass = "hep-ph",
    doi = "10.1007/JHEP11(2023)109",
    journal = "JHEP",
    volume = "11",
    pages = "109",
    year = "2023"
}

@article{Law-Smith:2023czn,
    author = "Law-Smith, Jamie A. P. and Obied, Georges and Prabhu, Anirudh and Vafa, Cumrun",
    title = "{Astrophysical constraints on decaying dark gravitons}",
    eprint = "2307.11048",
    archivePrefix = "arXiv",
    primaryClass = "hep-ph",
    doi = "10.1007/JHEP06(2024)047",
    journal = "JHEP",
    volume = "06",
    pages = "047",
    year = "2024"
}

@article{DeWolfe:1999cp,
    author = "DeWolfe, O. and Freedman, D. Z. and Gubser, S. S. and Karch, A.",
    title = "{Modeling the fifth-dimension with scalars and gravity}",
    eprint = "hep-th/9909134",
    archivePrefix = "arXiv",
    reportNumber = "HUTP-99-A048, MIT-CTP-2903",
    doi = "10.1103/PhysRevD.62.046008",
    journal = "Phys. Rev. D",
    volume = "62",
    pages = "046008",
    year = "2000"
}

@article{deGiorgi:2021xvm,
    author = "de Giorgi, Arturo and Vogl, Stefan",
    title = "{Dark matter interacting via a massive spin-2 mediator in warped extra-dimensions}",
    eprint = "2105.06794",
    archivePrefix = "arXiv",
    primaryClass = "hep-ph",
    doi = "10.1007/JHEP11(2021)036",
    journal = "JHEP",
    volume = "11",
    pages = "036",
    year = "2021"
}

@article{ArkaniHamed:1998nn,
    author = "Arkani-Hamed, Nima and Dimopoulos, Savas and Dvali, Gia",
    title = "{Phenomenology, astrophysics and cosmology of theories with submillimeter dimensions and TeV scale quantum gravity}",
    eprint = "hep-ph/9807344",
    archivePrefix = "arXiv",
    doi = "10.1103/PhysRevD.59.086004",
    journal = "Phys. Rev. D",
    volume = "59",
    pages = "086004",
    year = "1999"
}

@article{Mohapatra:2003ah,
    author = "Mohapatra, Rabindra N. and Nussinov, Shmuel and Perez-Lorenzana, Abdel",
    title = "{Large extra dimensions and decaying KK recurrences}",
    eprint = "hep-ph/0308051",
    archivePrefix = "arXiv",
    doi = "10.1103/PhysRevD.68.116001",
    journal = "Phys. Rev. D",
    volume = "68",
    pages = "116001",
    year = "2003"
}

@article{Dienes:2011ja,
    author = "Dienes, Keith R. and Thomas, Brooks",
    title = "{Dynamical Dark Matter: I. Theoretical Overview}",
    eprint = "1106.4546",
    archivePrefix = "arXiv",
    primaryClass = "hep-ph",
    doi = "10.1103/PhysRevD.85.083523",
    journal = "Phys. Rev. D",
    volume = "85",
    pages = "083523",
    year = "2012"
}

@article{Slatyer:2016qyl,
    author = "Slatyer, Tracy R. and Wu, Chih-Liang",
    title = "{General Constraints on Dark Matter Decay from the Cosmic Microwave Background}",
    eprint = "1610.06933",
    archivePrefix = "arXiv",
    primaryClass = "astro-ph.CO",
    doi = "10.1103/PhysRevD.95.023010",
    journal = "Phys. Rev. D",
    volume = "95",
    pages = "023010",
    year = "2017"
}

@article{Randall:1999ee,
    author = "Randall, Lisa and Sundrum, Raman",
    title = "{A Large mass hierarchy from a small extra dimension}",
    eprint = "hep-ph/9905221",
    archivePrefix = "arXiv",
    doi = "10.1103/PhysRevLett.83.3370",
    journal = "Phys. Rev. Lett.",
    volume = "83",
    pages = "3370--3373",
    year = "1999"
}

@article{Randall:1999vf,
    author = "Randall, Lisa and Sundrum, Raman",
    title = "{An Alternative to compactification}",
    eprint = "hep-th/9906064",
    archivePrefix = "arXiv",
    reportNumber = "MIT-CTP-2874, PUPT-1867, BUHEP-99-13",
    doi = "10.1103/PhysRevLett.83.4690",
    journal = "Phys. Rev. Lett.",
    volume = "83",
    pages = "4690--4693",
    year = "1999"
}

@book{Wald:1984rg,
    author = "Wald, Robert M.",
    title = "{General Relativity}",
    doi = "10.7208/chicago/9780226870373.001.0001",
    publisher = "Chicago Univ. Pr.",
    address = "Chicago, USA",
    year = "1984"
}

@article{Giudice:2017fmj,
    author = "Giudice, Gian F. and Kats, Yevgeny and McCullough, Matthew and Torre, Riccardo and Urbano, Alfredo",
    title = "{Clockwork/linear dilaton: structure and phenomenology}",
    eprint = "1711.08437",
    archivePrefix = "arXiv",
    primaryClass = "hep-ph",
    reportNumber = "CERN-TH-2017-219",
    doi = "10.1007/JHEP06(2018)009",
    journal = "JHEP",
    volume = "06",
    pages = "009",
    year = "2018"
}

@article{Dienes:2011sa,
    author = "Dienes, Keith R. and Thomas, Brooks",
    title = "{Dynamical Dark Matter: II. An Explicit Model}",
    eprint = "1107.0721",
    archivePrefix = "arXiv",
    primaryClass = "hep-ph",
    doi = "10.1103/PhysRevD.85.083524",
    journal = "Phys. Rev. D",
    volume = "85",
    pages = "083524",
    year = "2012"
}

@article{Vafa:2024fpx,
    journal = "arXiv e-prints",
    author = "Vafa, Cumrun",
    title = "{Swamplandish Unification of the Dark Sector}",
    eprint = "2402.00981",
    archivePrefix = "arXiv",
    primaryClass = "hep-ph",
    month = "2",
    year = "2024"
}

@article{Obied:2023clp,
    author = "Obied, Georges and Dvorkin, Cora and Gonzalo, Eduardo and Vafa, Cumrun",
    title = "{Dark dimension and decaying dark matter gravitons}",
    eprint = "2311.05318",
    archivePrefix = "arXiv",
    primaryClass = "astro-ph.CO",
    doi = "10.1103/PhysRevD.109.063540",
    journal = "Phys. Rev. D",
    volume = "109",
    number = "6",
    pages = "063540",
    year = "2024"
}

@article{Im:2024kuw,
    author = "Im, Sang Hui and Jod{\l}owski, Krzysztof",
    title = "{Searches for power-law warped extra dimensions}",
    eprint = "2412.20913",
    archivePrefix = "arXiv",
    primaryClass = "hep-ph",
    reportNumber = "CTPU-PTC-24-41",
    doi = "10.1007/JHEP01(2026)019",
    journal = "JHEP",
    volume = "01",
    pages = "019",
    year = "2026"
}

@article{Starkman:2000dy,
    author = "Starkman, Glenn D. and Stojkovic, Dejan and Trodden, Mark",
    title = "{Large extra dimensions and cosmological problems}",
    eprint = "hep-th/0012226",
    archivePrefix = "arXiv",
    reportNumber = "CWRU-P12-00, SU-GP-00-12-1",
    doi = "10.1103/PhysRevD.63.103511",
    journal = "Phys. Rev. D",
    volume = "63",
    pages = "103511",
    year = "2001"
}

@article{Kolb:1983fm,
    author = "Kolb, Edward W. and Slansky, Richard",
    title = "{Dimensional Reduction in the Early Universe: Where Have the Massive Particles Gone?}",
    reportNumber = "FERMILAB-PUB-83-088-A",
    doi = "10.1016/0370-2693(84)90298-3",
    journal = "Phys. Lett. B",
    volume = "135",
    pages = "378",
    year = "1984"
}

@article{Kaluza:1921tu,
    author = "Kaluza, Th.",
    title = {{Zum Unit{\"a}tsproblem der Physik}},
    eprint = "1803.08616",
    archivePrefix = "arXiv",
    primaryClass = "physics.hist-ph",
    reportNumber = "HUPD-8401",
    doi = "10.1142/S0218271818700017",
    journal = "Sitzungsber. Preuss. Akad. Wiss. Berlin (Math. Phys. )",
    volume = "1921",
    pages = "966--972",
    year = "1921"
}

@article{Klein:1926tv,
    author = "Klein, Oskar",
    editor = "Taylor, J. C.",
    title = "{Quantum Theory and Five-Dimensional Theory of Relativity. (In German and English)}",
    doi = "10.1007/BF01397481",
    journal = "Z. Phys.",
    volume = "37",
    pages = "895--906",
    year = "1926"
}

@unpublished{Lee:toappear,
    author = "Lee, Vincent S. H. and Randall, Lisa and Riojas, Marcos",
    title = "{To appear}",
    note = "to appear",
    year = "2026"
}

@article{Chivukula:2024nzt,
    author = "Chivukula, R. Sekhar and Gill, Joshua A. and Mohan, Kirtimaan A. and Sanamyan, George and Sengupta, Dipan and Simmons, Elizabeth H. and Wang, Xing",
    title = "{Limits on Kaluza-Klein portal dark matter models}",
    eprint = "2411.02509",
    archivePrefix = "arXiv",
    primaryClass = "hep-ph",
    doi = "10.1103/PhysRevD.111.075030",
    journal = "Phys. Rev. D",
    volume = "111",
    number = "7",
    pages = "075030",
    year = "2025"
}

@article{Chivukula:2025pmk,
    journal = "arXiv e-prints",
    author = "Chivukula, R. Sekhar and Gill, Joshua A. and Goh, Kenn S. and Mohan, Kirtimaan A. and Sanamyan, George and Sengupta, Dipan and Simmons, Elizabeth H. and Wang, Xing",
    title = "{Radion Portal Freeze-Out Dark-Matter}",
    eprint = "2507.21218",
    archivePrefix = "arXiv",
    primaryClass = "hep-ph",
    month = "7",
    year = "2025"
}

@article{Donini:2025qrf,
    author = "Donini, Andrea and Folgado, Miguel G. and Mu{\~n}oz-Ovalle, Alejandro",
    title = "{Dark matter in a three-brane Randall-Sundrum scenario out of the evanescent limit}",
    eprint = "2509.04580",
    archivePrefix = "arXiv",
    primaryClass = "hep-ph",
    doi = "10.1007/JHEP02(2026)206",
    journal = "JHEP",
    volume = "02",
    pages = "206",
    year = "2026"
}

@article{deGiorgi:2026qjp,
    journal = "arXiv e-prints",
    author = "de Giorgi, Arturo and Marcoli, Matteo and Silvetti, Federico",
    title = "{LHC Constraints on Resonant Kaluza-Klein Gravitons}",
    eprint = "2607.12012",
    archivePrefix = "arXiv",
    primaryClass = "hep-ph",
    reportNumber = "IPPP/26/31",
    month = "7",
    year = "2026"
}

@article{Bonifacio:2019ioc,
    author = "Bonifacio, James and Hinterbichler, Kurt",
    title = "{Unitarization from Geometry}",
    eprint = "1910.04767",
    archivePrefix = "arXiv",
    primaryClass = "hep-th",
    doi = "10.1007/JHEP12(2019)165",
    journal = "JHEP",
    volume = "12",
    pages = "165",
    year = "2019"
}

@article{Cesarotti:2020uod,
    author        = "Cesarotti, Cari and Reece, Matthew and Strassler, Matthew J.",
    title         = "{Spheres to Jets: Tuning Event Shapes with 5d Simplified Models}",
    eprint        = "2009.08981",
    archivePrefix = "arXiv",
    primaryClass  = "hep-ph",
    doi           = "10.1007/JHEP05(2021)096",
    journal       = "JHEP",
    volume        = "05",
    pages         = "096",
    year          = "2021"
}
\end{document}